\documentclass[aps,pre,floats,floatfix,twocolumn,fleqn]{revtex4-2}

\usepackage{graphicx}
\usepackage{epstopdf}
\graphicspath{{./Figs/}{./}} 

\usepackage{latexsym}
\usepackage{amsmath}
\usepackage{amssymb}
\usepackage{bm}
\usepackage{wasysym}
\usepackage{ifthen}
\usepackage{color}
\usepackage[colorlinks,allcolors=blue,bookmarks=true]{hyperref}

\newcommand{\trc}{\mbox{trace}}

\newcommand{\im}{\mbox{Im}}
\newcommand{\re}{\mbox{Re}}

\newcommand{\bra}[1]{\left\langle #1 \right|}
\newcommand{\ket}[1]{\left| #1 \right\rangle}
\newcommand{\braket}[1]{\left\langle #1 \right\rangle }

\newcommand{\kb}[2]{ | #1 \rangle \langle #2 |}

\newcommand{\beq}{\begin{eqnarray}}
\newcommand{\eeq}{\end{eqnarray}}

\newcommand{\hide}[1]{}  %{{\textcolor{red}{[hide]}}}
\newcommand{\rmrk}[1]{{#1}}

\newcommand{\Eq}[1]{{\textcolor{blue}{Eq.}}~\!\!(\ref{#1})} 

\newcommand{\Fig}[1] {{\textcolor{blue}{Fig.}}~\!\!\ref{#1}}
\newcommand{\sect}[1]{{\bf #1.-- }}

\newcommand{\hrefl}[2]{\href{#2}{(#1)}}

\begin{document}

% Lindblad spectrum of Anderson and Sinai disordered chains
% Research Proposal: Signature of localization in the Lindblad spectrum for Anderson-Sinai disorder
\title{Quantum walk in stochastic environment} 

\author{Ben Avnit, Doron Cohen} 

\affiliation{
\mbox{Department of Physics, Ben-Gurion University of the Negev, Beer-Sheva 84105, Israel} 
}

\begin{abstract}
We consider a quantized version of the Sinai-Derrida model for ``random walk in random environment". The model is defined in terms of a Lindblad master equation. For a ring geometry (a chain with periodic boundary condition) it features a delocalization-transition as the bias in increased beyond a critical value, indicating that the relaxation becomes under-damped. Counter intuitively, the effective disorder is enhanced due to coherent hopping. We analyze in detail this enhancement and its dependence on the model parameters. The non-monotonic dependence of the Lindbladian spectrum on the rate of the coherent transitions is highlighted.  
\end{abstract}

\maketitle

%\onecolumngrid

%%%%%%%%%%%%%%%%%%%%%%%%%%%%%%%%%%%%%%%%%%%%%%%%%%%%%%%%%%%%%%%%%%%%%%%%%%%%%%%%%%%%%%%%%%%%%%%%%%%%%%%%%%%%%%%%%%%%
\section{Introduction}

Sinai has coined the term "random walk in random environment" for a model that describes the stochastic motion of a particle in a 1D lattice \cite{Sinai1983}. The forward and backward rates $w_x^{\pm}$ of the transitions between sites (indexed by $x$) are independent random variables.  For a biased chain the average ratio $w_x^{+}/w_x^{-}$ favors (say) the forward direction. It turns out that for an unbiased infinite chain with arbitrarily small randomness the spreading of the particle becomes sub-diffusive. Later Derrida and followers \cite{Derrida1983,BOUCHAUD1990,BOUCHAUD1990a} have found that non-zero drift velocity is induced if the bias exceeds a critical value, aka {\em sliding transition}.  Related to that is the {\em delocalization transition} that has been discussed by Hatano, Nelson and followers  \cite{Hatano1996,Hatano1997,Shnerb1998,Feinberg1999,KAFRI2004,KAFRI2005}. The latter term refers, in the Sinai-Derrida context, to the transition from over-damped to under-damped {\em relaxation} for a finite sample with periodic boundary conditions \cite{neg,nep,nts,rxa}.

We consider a quantum version of Sinai-Derrida model. This means that in addition to the stochastic transitions that are described by an appropriate master equation, the particle can also perform coherent hopping between the sites. The hopping frequency~$c$ is a free model parameter. Our interest is focused in the regime ${c\ll\nu}$, where $\nu$ is the average rate of the stochastic transitions. Note that in the other extreme ($\nu{=}0$) the model features ballistic motion that can be suppressed by an Anderson localization effect (due to quenched disorder), or by Bloch oscillations (if bias is applied). 

In \cite{qss} we have introduced a full Ohmic Lindbladian that generates the quantized version of the Sinai-Derrida model. A counter-intuitive enhancement of the effective disorder due to coherent hopping has been pointed out, but has not been explored. In particular, the most interesting aspect, namely, the delocalization transition, has not been discussed. In the present paper we consider a {\em minimal version} of the full quantized version, omitting some terms that are not essential for the demonstration of the main effects, and performing some further simplifications that will be discussed in subsequent sections. Thus, in the absence of coherent hopping ($c{=}0$) our minimal model reduces to the Pauli master equation, and hence becomes identical to the standard Sinai-Derrida model. 
%
%The full Ohmic model \cite{qss,qsd} includes some secondary quantum terms, and consequently features quantum signatures that are not related to the essential physics of the delocalization transition, that are ignored in the present work.          

The minimal model that we introduce below is defined by a Lindbladian. It includes a random potential that has dispersion $\sigma_{\mathcal{E}}$, and a random stochastic field that has dispersion $\sigma_f$. The parameters that define the model are ${ (\nu,c,\sigma_{\mathcal{E}},\sigma_f) }$ and the bias $f$.  We argue that such Lindbladian reflects an environment that has a characteristic temperature 
\beq \label{eT}
T_{\text{bath}} \ = \ (\sigma_f/\sigma_{\mathcal{E}})^{-1}
\eeq
An associated dimensionless parameter is 
\beq \label{eETA}
\eta \ = \ \frac{\nu}{2T_{\text{bath}}} \ = \ \frac{1}{2} \left(\frac{\sigma_f}{\sigma_{\mathcal{E}}}\right) \, \nu
\eeq
Accordingly, there are two ``classical'' dimensionless parameters
and two ``quantum" dimensionless parameters that define the model:
\beq
\text{Dimensionless Parameters} \ = \ (f, \sigma_f, \eta, c/\nu) 
\eeq

\begin{figure}[h!]
\centering
\includegraphics[width=5cm]{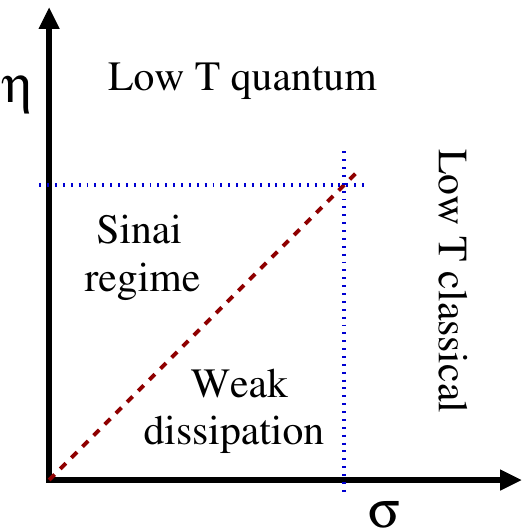}
\caption{
{\bf Regime diagram.}    
The $(\sigma_f, \eta)$ regime diagram for the unbiased model. 
The classical high temperature condition (${\sigma_f<1}$) implies weak stochastic field. 
The quantum high temperature condition (${\eta < 1}$) allows to ignore memory effects.
In the Sinai regime, above the dashed line (${\eta > \sigma_f}$), it is allowed to regard  
the coherent effects as a perturbation with respect to the dominant stochastic dynamics.  
}
\label{fRegimes}
\end{figure}

\sect{Outline}
We introduce the stochastic and the quantized models in Sections~II and~III, with extra technical details in Appendix~A. We further discuss the significance of the model parameters in section~IV, and provide a regime diagram in \Fig{fRegimes}. Then we look on the spectrum of the Lindbladian for non-disordered and for disordered ring in Sections~V and~VI respectively.  We discuss how the localization of its eigen-modes is affected by~$c$ in Section~VII, and highlight some counter-intuitive effects. The delocalization threshold is further explained in Section~VIII. The summary in Section~VII provides extra background, to place the present work in the context of past studies.

%%%%%%%%%%%%%%%%%%%%%%%%%%%%%%
%%%%%%%%%%%%%%%%%%%%%%%%%%%%%%

%%%%%%%%%%%%%%%%%%%%%%%%%%%%%%%%%%%%%%%%%%%%%%%
\section{The Stochastic model}
% introducing W

The standard Sinai-Derrida model is defined in terms of a rate equation for the probabilities $p_x$ to find the particle in site ${x=1,2,...N}$, and we assume periodic boundary conditions. The rate equation is written as follows,   
\beq \label{eW}
\frac{d}{dt} \bm{p} \ = \ \mathcal{W} \bm{p} 
\eeq  
where $\bm{p} = \{p_x \}$ is a vector, and $\mathcal{W}$ is an $N\times N$ matrix. 
The explicit expression for this matrix is 
\beq
\mathcal{W} = -\sum_{x} (w^{+}_x+w^{-}_{x{-}1}) \bm{Q}_{x} 
+ \sum_{x} \left[ w^{+}_x \bm{D}_{x} +  w^{-}_x \bm{D}_{x}^{\dag} \right]  
\ \ \ \ \ 
\eeq    
where $\bm{Q}_x = \kb{x}{x}$ and $\bm{D}_x = \kb{x+1}{x}$. 
The translation operator is ${\bm{D} = \sum_x \bm{D}_x = e^{-i\bm{q}} }$,
where $\bm{q}$ is the generator of translations.  
In the absence of disorder the above expression takes the form
\beq
\mathcal{W} = - (w^{+}+w^{-})[1-\cos(\bm{q})] - i (w^{+}-w^{-}) \sin(\bm{q}) 
\ \ \ \ \
\eeq 

The rates $w_x^{\pm}$ in the disordered Sinai model   
are determined by a random stochastic field $f_x$ such that 
\beq \label{efxDef}
\frac{w^{-}_x}{w^{+}_x} \ \equiv \ e^{-f_x}  
\eeq
We assume, following Sinai, weak stochastic disordered (${ f_x \ll 1 }$).
Consequently, one writes in leading order 
\beq \label{efx}
w^{\pm}_x \ \equiv \  \nu_x e^{\pm f_x/2}  \ \approx \ \left( 1 \pm \frac{f_x}{2} \right) \nu_x 
\eeq 
Accordingly, $\nu_x$ characterizes the strength of the stochastic transitions 
at a given bond, while $f_x$ reflects their asymmetry.

For the later analysis we write an explicit expression for the $\mathcal{W}$ matrix,
that holds in leading order with respect to the disorder strength:  
\beq 
\mathcal{W} &=& -\text{diagonal}\left\{ (\nu_{x}{+}\nu_{x{-}1}) + \frac{\nu}{2}(f_{x}{-}f_{x{-}1}) + \frac{\nu}{4} f^2 \right\} 
\nonumber \\ \label{eWmatrix}
&& + \text{offdiagonal}\left\{ \nu_x e^{\pm f_x/2} \right\}  
\eeq 
In the above formula the off diagonal terms are written without any approximation, 
because it is more convenient for later discussion. 
But a clarification is required for the approximations that are involved in the diagonal terms. 
The term ${ \nu (f_{x}-f_{x{-}1}) }$ 
is implied by the replacement of ${\nu_x}$ by its average value $\nu$. 
The error that is associated with this replacement is of higher order in the disorder strength. 
The same reasoning applies for the $\nu f^2$ term, 
where the replacement of $f_x$ by its average value $f$ has been performed.

%%%%%%%%%%%%%%%%%%%%%%%%%%%%%%
\begin{figure}
\centering
\includegraphics[width=5cm]{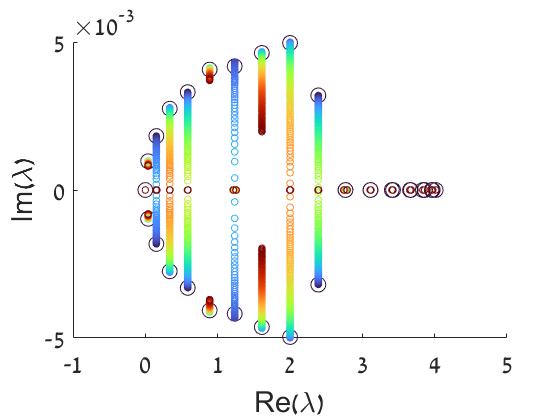} 
\includegraphics[width=5cm]{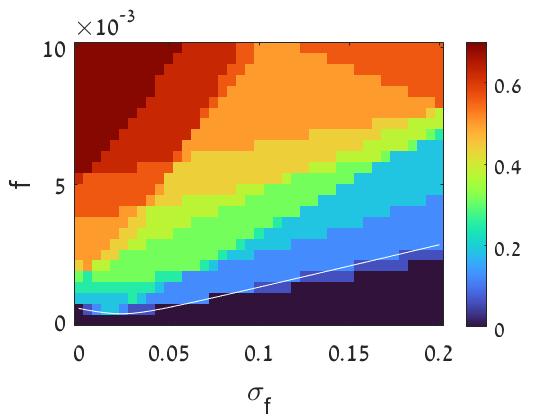} 
\caption{
{\bf Delocalization of the eigen-modes.}    
The chain consist of $N=32$ sites with periodic boundary conditions. 
The dynamics is described by a rate equation with matrix $\mathcal{W}$.  
The average transition rate between neighboring sites is ${\nu=1}$, 
and \rmrk{${ \sigma_{\nu}=0.05 }$}. 
{\em Upper panel:}~The spectra of $\mathcal{W}$ for \rmrk{${f=0.005}$}. 
The color code indicates the value of~$\sigma$.
It goes from blue ($\sigma_f{=}0$) to red \rmrk{($\sigma_f{=}0.02$)}.  
{\em Lower panel:}~Characterization of the spectrum in a wider range.
The axes are ${(\sigma_f,f)}$. 
The color-code indicates the normalized number $N_{\text{cmplx}}/N$ of complex eigenvalues.
The solid line indicates the numerically determined critical value $f_c$ 
above which the eigenvalues at the vicinity of $\lambda{=}0$ become complex. 
For different realizations of the disorder this line is shifted within some range. 
}
\label{fNcmplx}
\end{figure}
%%%%%%%%%%%%%%%%%%%%%%%%%%%%%%

The random independent variables $f_x$ are characterized by  
an average $f_{\text{bias}} \equiv f$, and by a dispersion $\sigma_f$.     
The spectrum of $\mathcal{W}$ is illustrated in \Fig{fNcmplx}. 
As $f$ is increased more eigenvalues become complex (see lower panel). 
The critical value $f_c$ is the value above which complex eigenvalues 
emerge at the vicinity of ${\lambda{\sim}0}$.
This is identified as a {\em delocalization transition} in the sense of Hatano and Nelson,  
and has a subtle relation \cite{neg} to the {\em sliding transition} 
that has been discussed by Derrida and followers.  
An estimate for $f_c$ can be obtained by the formula
\beq \label{efc}
f_c \ = \ \frac{1}{4}\text{Var}(f_x) \ = \ \text{prefactor} \ \sigma_f^2
\eeq
where the numerical prefactor depends on the 
numerical definition of $\sigma_f$ that may vary 
depending on the shape (Gaussian / Box) of the distribution. 
This expression works well for a long chain, 
while fluctuations in its value are pronounced for short samples.

%%%%%%%%%%%%%%%%%%%%%%%%%%%%%%%%%%%%%%%%%%%%%%%
\section{The Quantized model}
% introducing L

The full Ohmic version of the Lindblad equation for an $N$ site chain with periodic boundary conditions can be found in Appendix~A. Here we summarize the details of a simplified minimal version that still contains all the essential physics of the problem under study. The master equation for the evolution of the probability matrix is  
\beq\label{eqL}
\frac{d\rho}{dt} = \left(\mathcal{L}^{(\mathcal{H})} + \mathcal{L}^{(bias)} + \mathcal{L}^{(B)} + \mathcal{L}^{(S)} \right) \rho 
\eeq
The Lindblad generators in this equation refer to the coherent Hamiltonian dynamics, to the coherent bias term, to the stochastic environmentally-induced transitions between sites (along ``$B$onds"), and to optional decoherence due to local baths (at ``$S$ites"). The explicit expressions are:       
\beq
\mathcal{L}^{(\mathcal{H})} \rho =&&  -i[\bm{H},\rho] \\
\mathcal{L}^{(B)}\rho =&&   
-\frac{1}{2}\sum_{x} (w^{+}_x+w^{-}_{x{-}1})\left[ \bm{Q}_{x} \rho + \rho \bm{Q}_{x} \right]
\nonumber \\
&& 
+ \sum_{x} \left[ w^{+}_x \bm{D}_{x}^{\dag} \rho \bm{D}_{x} +  w^{-}_x \bm{D}_{x} \rho \bm{D}_{x}^{\dag} \right]   \\
\mathcal{L}^{(S)}\rho =&&  -\gamma \rho + \sum_{x} \gamma  \bm{Q}_{x} \rho \bm{Q}_{x}  
%&& \mathcal{L}^{(B)}\rho = \sum_{x} \nu_{x} \left[ \bm{L}_{x} \rho \bm{L}_{x}^{\dagger} \ - \frac{1}{2} \{ \bm{L}_{x}^{\dagger} \bm{L}_{x}, \rho \}  \right] \\
%&& \mathcal{L}^{(S)}\rho = -\gamma \left[ \rho - \sum_{x} \bm{Q}_{x} \rho \bm{Q}_{x} \right]
\eeq
The stochastic transition rates are $w^{\pm}_x$ as in \Eq{efx}, 
and the extra on-site decoherence rate is~$\gamma$. 
The Hamiltonian incorporates a hopping term and a disordered potential: 
\beq
\mathcal{H} \ = \ \frac{c}{2}(\bm{D} + \bm{D}^{\dagger}) + U(\bm{x})  
% \ = \  c\cos(\mathbf{p}) + U(\bm{x})
\eeq
where $c$ is the hopping frequency for coherent transitions. 
The disordered field is 
\beq
\mathcal{E}_x \ = \ -(U(x{+}1) - U(x))
\eeq
If we did not impose periodic boundary conditions, 
the bias could have been added using the prescription 
${ U(x) \mapsto U(x) - \mathcal{E} x }$, 
with diagonal matrix elements ${ i (x_n-x_m) \mathcal{E}_{bias} }$. 
In order to respect the periodic boundary condition we modify 
the bias term as follows:
\beq 
&& \mathcal{L}^{(bias)}(n',m'|n,m)  = 
\nonumber \\  \label{eq:LnoBias}
&& \ \ \ \ \ \ 
i \delta_{n',n}\delta_{m',m} \ \mathcal{E}_{bias} \frac{N}{2 \pi} \sin{\Big(\frac{2\pi}{N}(x_n-x_m)\Big)}  
\ \ \ \ \ \ \ \ \ \ \ \ 
\eeq
This modified version is locally the same as the proper version for an open chain, 
while for large $(x_n-x_m)$ it can be justified self-consistently for a long closed chain.  
This modification has no significant numerical implication, because the far off-diagonal terms of the probability matrix for low modes are vanishingly small.

The definition of the local temperature $T_x$ 
is implied by the Boltzmann ratio \Eq{efxDef},
using the substitution 
\beq \label{efx}
f_x  \ \equiv \ \frac{\mathcal{E}_x}{T_x}
\eeq
Recall that we assume, following Sinai, weak stochastic disordered (${ f_x \ll 1 }$), 
which is equivalent to ${ \mathcal{E}_x \ll T_x }$. 
This goes well with the observation that the Ohmic approximation is consistent 
with Boltzmann to leading order in~$1/T$ 
(higher order terms in the Ohmic master equation vanish only in the classical limit). 
In Appendix~A we explain how \Eq{efx} is obtained rigorously 
from the Ohmic master equation. The free parameters of the 
the Ohmic master equation are $\nu_x$ and $\eta_x$ that correspond to 
the "noise" intensity and the "friction" coefficient in the common Langevin description.
They obey the Einstein relation, namely, ${T_x = \nu_x/(2\eta_x)}$. 
However, in the present model the coupling of the bonds to the baths implies 
that~$\eta$ should be regraded as a "mobility" and not as "friction" coefficient \cite{qss}.

%%%%%%%%%%%%%%%%%%%%%%%%%%%%%%%%%%%%%%%%%%%%%%%%%%%%%%%%%%%%%%%%%%%%%%%%%%%%%%%%%%%%%%%%%%%%%%%%%%%%%%%%%%%%%%%%%%%%
\section{Model parameters}

Physically disorder may arise from the potential, or from the environmental parameters. So we may have randomness in $\nu_x$ and/or in $\eta_x$ and/or in $\mathcal{E}_x$ and/or in $\gamma_x$. The Sinai-Derrida physics that we discuss is rather robust and allows flexibility in the choice of the ``free" parameters. In the numerical study, the following approach has been adopted with no loss of generality. Given $\sigma_{\mathcal{E}}$ we generate a realizations of the disordered potential such that 
\beq \label{eSU}
U(x) \in [0,\sigma_{\mathcal{E}}] 
\eeq
Then we can generate a random ${\eta_x \in [0, \sigma_{\eta}]}$, 
and from it to calculate the random stochastic field $f_x$. 
In practice we have realized that the numerical results are robust, 
and not affected if we generate the stochastic field independently, namely,  
\beq \label{eSf}
f_x \in \left[f-\frac{\sigma_f}{2}, f+\frac{\sigma_f} {2} \right]
\eeq
with  
\beq
\text{Var}(f_x)  \ = \ \frac{C}{\nu^2} \sigma_{\eta}^2  \ \text{Var}(\mathcal{E}_x)  
\eeq
The latter relation follows from Eq(24) of \cite{nts}, 
where $C=8$ for Gaussian disorder. 
From this relation it follows that the ratio $\sigma_f/\sigma_{\mathcal{E}}$ 
is determined by the temperature of the bath. 
This inspires the practical definition of the characteristic temperature \Eq{eT}.

%%%%%%%%%%%%%%%%%%%%%%%%%%%%%%%%%%%%%%
\sect{Resistor network disorder} 
The essential type of disorder for the discussion of Sinai-Derrida Physics is related to the randomness of the stochastic field $f_x$. As opposed to that, randomness in $\nu_x$ is similar to ``resistor network disorder". It has significant implications only in extreme circumstances, such that {\em percolation} becomes an issue \cite{neg}. We assume weak disorder, and therefore the probability for disconnected bonds is zero. For the numerical exploration we take 
\beq \label{eSnu}
\nu_x \ \ \in \ \ \left[ \nu -\frac{\sigma_{\nu}}{2}, \ \nu+\frac{\sigma_{\nu}}{2} \right]
\eeq
where $\nu$ is the average value of $\nu_x$, 
and ${ \sigma_{\nu} \ll \nu  }$ is assumed.
%
% Unless stated otherwise we assume ${\sigma_{\nu}=0}$. However, in a later stage of our analysis, we are going to argue that a $c$~dependent contribution to ${\sigma_{\nu}$ is implied if we reduce the quantized model into an effective stochastic model.   

%%%%%%%%%%%%%%%%%%%%%%%%%%%%
\sect{Numerical procedure} 
Given $\sigma_{\nu}$ we generate random set of $\nu_x$ values for the bonds in accordance with \Eq{eSnu}. 
We set the units of time such that the average value is ${\nu=1}$. 
Given $\sigma_{f}$ and $f$, we generate random realizations of the stochastic field $f_x$ 
in accordance with \Eq{eSf}, such that $\sum f_x \equiv Nf$ for each realization.
Note that the average value ${f}$, per realization, 
is regarded as a control parameter, namely ${f_{\text{bias}} \equiv f}$.
The transition rates $w_x^{\pm}$ are calculated using \Eq{efx}. 
Given $T_{\text{bath}}$ we determine $\sigma_{\mathcal{E}}$ from \Eq{eT}, 
and generate a realization of the disordered potential in accordance with \Eq{eSU}.            
Respectively in \Eq{eq:LnoBias} we substitute 
${ \mathcal{E}_{\text{bias}} = T_{\text{bath}} f_{\text{bias}} }$.

%%%%%%%%%%%%%%%%%%%%%%%%%
\sect{Regime diagram} 
The $(\sigma_f, \eta)$ regime diagram of the unbiased model is displayed in \Fig{fRegimes}. 
The assumed hierarchy of energy scales is 
\beq \label{eHR}
c, \ \sigma_{\mathcal{E}} \ \ \ll \ \ \nu \ \ \ll \ \ T_{\text{bath}}
\eeq 
The horizontal axis of the diagram is the strength $\sigma_f$ of the Sinai disorder, 
that is determined by the ratio between $\sigma_{\mathcal{E}}$ and $T_{\text{bath}}$ via \Eq{eT}. 
As already pointed out we assume weak stochastic field ($\sigma_f \ll 1$) 
reflecting that we deal with an Ohmic master equation that corresponds 
to the traditional Sinai-Derrida model.
With similar reasoning we assume ${ c \ll T }$ for the coherent hopping.    
The vertical axis of the diagram is the quantum parameter $\eta$ that reflects the ratio 
between $\nu$ and $T_{\text{bath}}$. It is the same dimensionless ``friction" parameter that appears is the analysis 
of the Spin-Boson Hamiltonian. The validity of the Ohmic master equation requires ${\eta \ll 1}$. In the regime ${\eta > 1}$ the model is not valid because non-Markovian memory effects cannot be neglected. 

The first inequality in \Eq{eHR} means that we regard the coherent effects as a perturbation with respect to the dominant stochastic dynamics. This stands in opposition to the common quantum-dissipation studies, where the bath is regraded as a disturbance that slightly spoils or modifies coherent evolution. In the regime diagram the  border between the two regimes is represented by the diagonal line ${\eta=\sigma_f}$.  

In our mathematical analysis $T_{\text{bath}}$ merely determines the 
ratio $\sigma_{\mathcal{E}}/\sigma_f$, 
and should be kept larger than $\nu \equiv 1$ in accordance with \Eq{eHR}.    
To avoid misunderstanding, we emphasize that from an experimental perspective the physical temperature affects the parameters $\nu$ and $\sigma_f$. Therefore, setting ${\eta = \infty}$ in the sense of \Eq{eETA}, while {\em fixing} the transition rates, does not really corresponds to zero temperature, and furthermore contradicts our assumption \Eq{eHR}.

%%%%%%%%%%%%%%%%%%%%%%%%%%%%%%%%%%%%%%%%%%%%%%%%%%%%%%%%%%%%%%%%%%%%%%%%%%%%%%%%%%%%%%%%%%%%%%%%%%%%%%%%%%%%%%%%%%%%
\section{The non-disordered ring}
% including Pauli for c=0

For non-disordered ring with ${c=0}$ the Lindblad equation becomes identical with the Pauli master equation, namely, 
the diagonal elements $p_x$ of the probability matrix $\rho_{x',x''}$ satisfy the rate equation \Eq{eW}, while each diagonal term satisfies the equation 
\beq
\frac{d}{dt}\rho_r \ = \  [-\gamma -(w^{+}+w^{-})  + i \mathcal{E}r] \rho_r   
\eeq   
with ${r=(x'{-}x'')=\pm1,\pm2,\cdots }$.   
We conclude that the eigenvalues $\{ -\lambda_{q,r} \}$  of $\mathcal{L}$ are 
\beq
\lambda_{q,r{=}0} &=& (w^{+}{+}w^{-})[1{-}\cos(q)] - i(w^{+}{-}w^{-})\sin(q)  
\ \ \ \ \ \ \ \ \ \ \\ \label{eSpecST}
\lambda_{q,r{\ne}0} &=& \gamma + (w^{+}{+}w^{-}) - i \mathcal{E} r
\eeq   
where the wavenumber is ${q=(2\pi/N)\times\text{integer}}$.
Accordingly, we distinguish between {\em relaxation-modes} that have eigenvalues $\lambda_{q,0}$, and {\em decoherence-modes} that have eigenvalues $\lambda_{q,r{\ne}0}$. 
This distinction is blurred for ${c \ne 0}$ due to mixing of the $r$ branches, 
but nevertheless it can be maintained for small~$c$ (see below), 
even in the presence of disorder (see next Section).

We can extract the drift velocity $v$, and the diffusion coefficient $D$   
from the expansion 
\beq
\lambda_{q,0} \ \approx \ ivq + Dq^2 + \mathcal{O}(q^3)
\eeq
For non-disordered ${c=0}$ ring, \Eq{eSpecST} implies, as expected, the trivial results  
${v=(w^{+}-w^{-})}$ and ${D=(w^{+}+w^{-})/2}$.

\begin{figure}
\centering
\includegraphics[width=6cm]{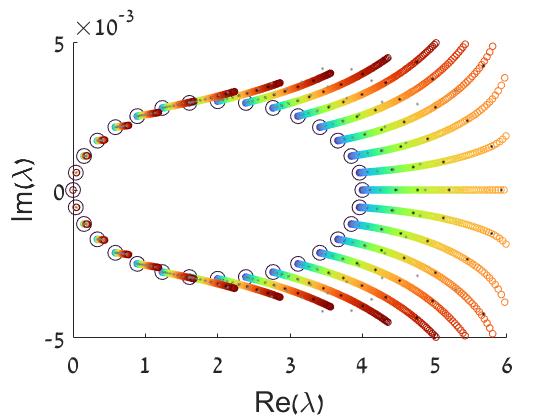}
\includegraphics[width=6cm]{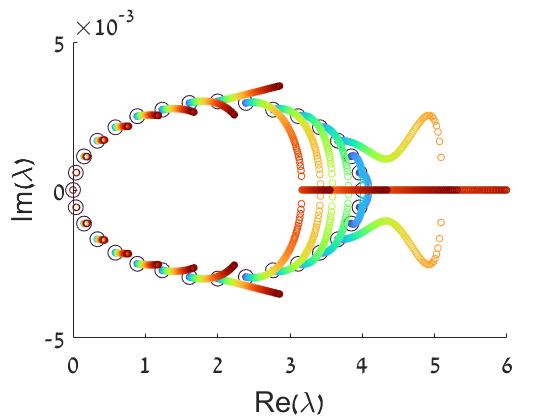}
\includegraphics[width=6cm]{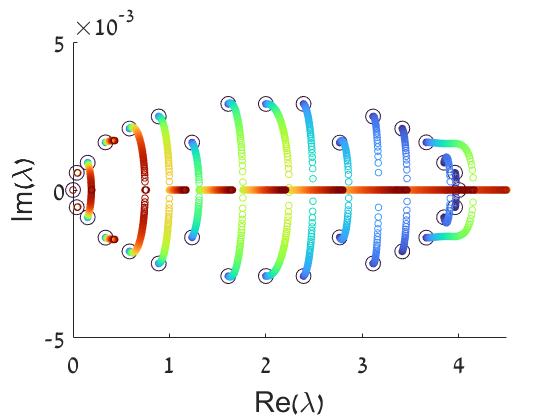}
\caption{
{\bf The Lindblad spectrum.}    
Spectra of the Lindbladian for the same ring 
as in \Fig{fNcmplx}, with ${\sigma_{\nu}=0}$, 
while the dispersion of the stochastic field 
is ${\sigma_f=0}$ (upper panel), 
and ${\sigma_f=0.005}$ (middle panel),
and ${\sigma_f=0.02}$ (lower panel). 
The color code indicates the value of~$c$. 
It goes from blue ($c{=}0$) to red ($c{=}2$). 
The bias is \rmrk{ ${f=0.003}$ }. 
The temperature is \rmrk{${ T_{\text{bath}}=200 }$ }, 
and the on-site decoherence rate is \rmrk{${\gamma =5}$}.
The black circles indicate the spectrum of $\mathcal{W}$.
In the upper panel the black dots are from \Eq{eSpec}, 
while the gray dots are from the perturbative approximation based on \Eq{eLqt}.
The eigenvalues that correspond to the decoherence-modes \Eq{eSpecST}
are outside of axis borders. 
}
\label{fSpctraND}
\end{figure}

Next we would like to explore how the spectrum is modified for ${c \ne 0}$. 
An example for the outcome of numerical diagonalization is provided in \Fig{fSpctraND}. 
The dependence on $c$ is illustrated.
Below we explain the observed dependence analytically.

%%%%%%%%%%%%%%%%%%%%%%%%%%%%%%
%%%%%%%%%%%%%%%%%%%%%%%%%%%%%%

%%%%%%%%%%%%%%%%%%%%%%%%%%%%%%
\begin{figure}
\centering
\includegraphics[width=5cm]{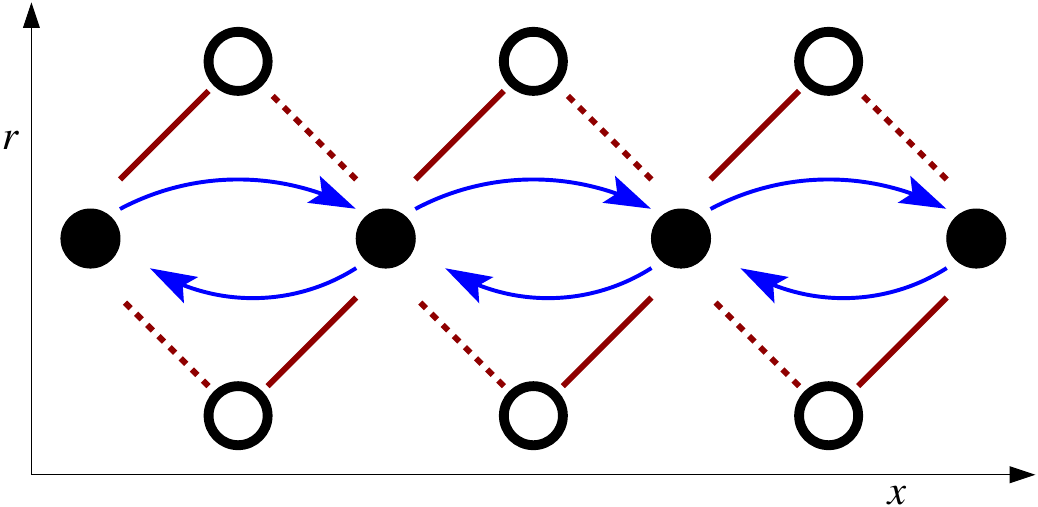}
\caption{
{\bf Diagrammatic representation of the lattice.}
The axes are $(x,r)$. Only the ${r=-1,0,1}$ elements are displayed.  
Bath-induced $w^{\pm}$  transitions are colored in blue:  
they connect only ${r=0}$ elements.  
Coherent $\pm i(c/2)$ transverse transitions are indicated by solid and dashed red lines. 
}
\label{fBloch}
\end{figure}
%%%%%%%%%%%%%%%%%%%%%%%%%%%%%%

The hopping couples the diagonal and the off diagonal terms of $\rho_{x',x''}$. 
It is convenient to define a position coordinate 
${x=(x'{+}x'')/2}$ and a transverse coordinate ${r=(x'{-}x'')}$.  
Then we can define an operator $\bm{r}=\sum_r \ket{r}r\bra{r}$, 
and a displacement  operator  $\mathcal{D}_{\perp}$ is the transverse $r$ coordinate, 
and a displacement  operator $e^{-i\bm{q}}$ in the $x$ coordinate.    
The total Lindbladian can be regarded as a non-Hermitian Hamiltonian 
that generates dynamics on an $(x,r)$ lattice, see \Fig{fBloch}. 
It can be expressed as follows:
\beq
\mathcal{L} &=& - \gamma_0  \ +  (\gamma_0 -\mathcal{W}) \otimes \kb{0}{0}  
\nonumber \\
&& -i \mathcal{E} \bm{r} 
\ - c \sin(\bm{q}/2) \left[\mathcal{D}_{\perp} - \mathcal{D}_{\perp}^{\dag}\right]
\eeq 
where ${\gamma_0 \equiv  \gamma + w^{+} + w^{-} }$. 
More generally we define ${\gamma_q \equiv  \gamma + w^{+} e^{-iq} + w^{-} e^{iq}}$.
In the absence of disorder the lattice has Bloch translation symmetry in $x$, 
and therefore $q$ is a good quantum number. The $q$ block of the Lindbladian is 
\beq
\mathcal{L}^{(q)} = - \gamma_0  +  \gamma_q  \kb{0}{0} 
-i \mathcal{E} \bm{r} 
- c \sin(q/2) \left[\mathcal{D}_{\perp} - \mathcal{D}_{\perp}^{\dag}\right]
\ \ \ \ \ \ 
\eeq
For clarity, and for further analysis, 
we write a truncated matrix version of  $\mathcal{L}^{(q)}$,  
where we keep only ${r=-1,0,1}$. Namely, 
\beq \label{eLqt} 
\mathcal{L}^{(q)}  = 
\begin{pmatrix}
-\gamma_0 + i\mathcal{E} & c\sin(q/2)       & 0 \\
-c\sin(q/2)    &  -\gamma_0 + \gamma_q       & c\sin(q/2) \\
0             & -c\sin(q/2) & -\gamma_0 - i\mathcal{E}
\end{pmatrix}
\ \ \ \ \ \ \ 
\eeq   
In section~4 of the supplementary of \cite{qss} (see also \cite{EspositoGaspard2005}),
the following result has been derived:
\beq  \label{eSpec}
\lambda_{q,0} \ \ = \ \ \gamma_0 - \sqrt{\gamma_q^2 - 4 c^2 \sin^2{(q/2)}} 
\eeq
This result allows finite $f_{\text{bias}}$ 
but neglects $\mathcal{E}_{\text{bias}}$.

The eigenvalues of $\mathcal{L}^{(q)}$ are labeled $\lambda_{q,s}$, 
with band index ${s=0,\pm1,\pm2,\cdots}$ that distinguishes the ${s=0}$ relaxation modes, 
from the ${s\ne0}$ decoherence modes. 
The former correspond to the eigenvalues of $\mathcal{W}$. 
The distinction between {\em  relaxation modes} 
and  {\em decoherence modes} remains meaningful for small $c$, 
as long as the bands remain separated.    
In \Fig{fSpctraND} only the eigenvalues of the relaxation modes are displayed. 
As $c$ becomes larger, the $\im(\lambda)$ of the relaxation modes increases monotonically. 
The numerical diagonalization agree with \Eq{eSpec}, 
and approximately with diagonalization  of the truncated version \Eq{eLqt}.

%%%%%%%%%%%%%%%%%%%%%%%%%%%%%%%%%%%%%%%%%%%%%%%%%%%%%%%%%%%%%%%%%%%%%%%%%%%%%%%%%%%%%%%%%%%%%%%%%%%%%%%%%%%%%%%%%%%%
\section{The effect of disorder}

The matrix $\mathcal{W}$ is real. Therefore its characteristic polynomial is real, and accordingly its eigenvalues $\lambda_q$
are either real or complex-conjugate pairs.  Note that in the absence of disorder $q$ can be interpreted as quasi-momentum, while in
the presence of disorder $q$ becomes a dummy index.    
The delocalization of eigen-modes, as $f_{\text{bias}}$ in increased, is indicated by the formation of complex-conjugate pairs. 
The eigenvalues at the vicinity of ${\lambda \sim 0}$ are the first to get delocalized, indicating a crossover from over-damped to under-damped relaxation \cite{neg,nep}. For very large $f_{\text{bias}}$ most of the eigenvalues, also those with large $\re(\lambda)$, become complex. \Fig{fNcmplx} illustrates this delocalization scenario, showing how the number of complex eigenvalues depends on $f$ for a range of $\sigma_f$ values. 

Similar scenario is expected for the Lindbladian $\mathcal{L}$. The hermiticiy of $\rho$ implies that the super-matrix $\mathcal{L}_{n',m'|n'',m''}$ is complex conjugated if we perform the reflection ${\mathcal{R}: (n,m) \mapsto (m,n)}$. So we have the relation ${\mathcal{R} \mathcal{L}\mathcal{R} = \mathcal{L}^* }$.  This implies that the characteristic 
polynomial of $\mathcal{L}$ is real, as in the case of $\mathcal{W}$.

%%%%%%%%%%%%%%%%%%%%%%%%%%%%%%
\begin{figure}
\centering
\includegraphics[width=4cm]{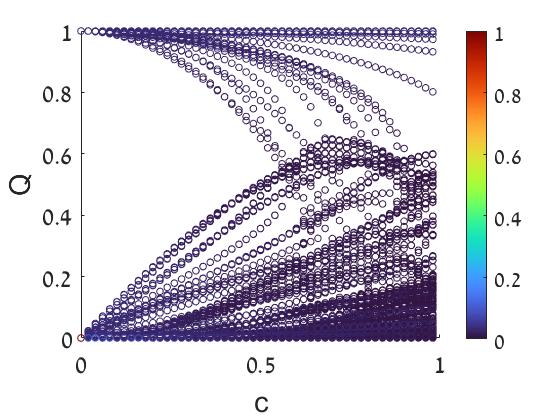}
\includegraphics[width=4cm]{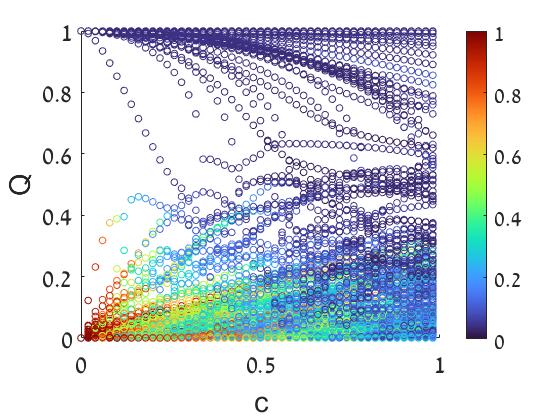}\\
\includegraphics[width=4cm]{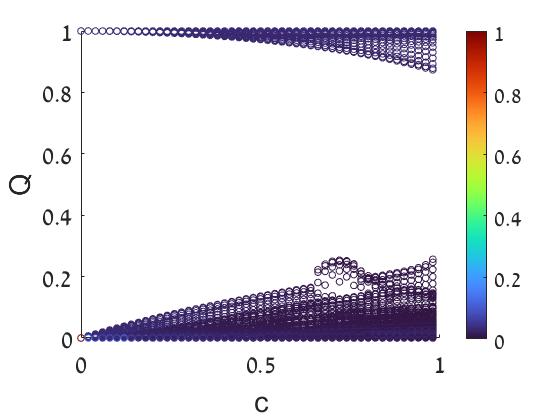}
\includegraphics[width=4cm]{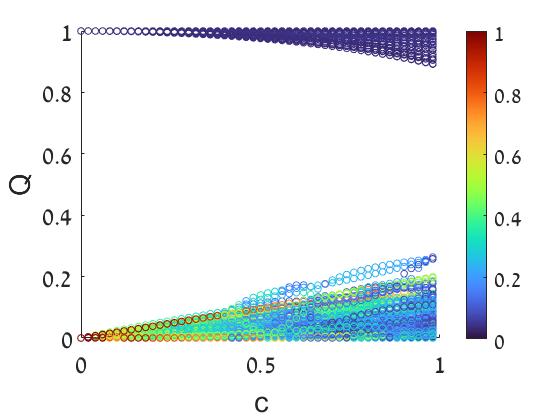}
\caption{
{\bf The diagonal norm versus $c$.}    
We consider the same ring as in \Fig{fNcmplx} 
with \rmrk{$\gamma{=}0$} (upper panels) and \rmrk{$\gamma{=}5$} (lower panels) .  
For a given value of $c$ we calculate the diagonal norm $Q$ 
and the IPR (color coded) for each eigen-mode.    
The measure $Q$ allows to distinguish the ${s=0}$ branch from the other branches. 
As $c$ is increased the branches get-mixed.
The left panels are for zero disorder, 
while the right panels are for \rmrk{${\sigma_f=0.02}$}. 
}
\label{fdiagNorm}
\end{figure}
%%%%%%%%%%%%%%%%%%%%%%%%%%%%%%

%%%%%%%%%%%%%%%%%%%%%%%%%%%%%%%%%%%%%
\sect{The relaxation spectrum}
The relaxation spectrum of a disordered ring for $c{\neq}0$ is illustrated 
in \Fig{fSpctraND}. Note that we use the same ring as in \Fig{fNcmplx}, 
with the same disorder realization. 
Different values of $\sigma_f$ are achieved by uniform ``stretching" of the field values, 
without affecting the relative magnitudes.     
The major counter-intuitive observation is as follows: the introduction of coherent hopping 
is qualitatively similar to {\em stronger} disorder. This is reflected by the migration of eigenvalues towards the real axis. The effect is pronounced for eigenvalues with larger $\re(\lambda)$, namely, eigenvalues with larger $\re(\lambda)$ are more sensitive to~$c$.

%%%%%%%%%%%%%%%%%%%%%%%%%%%%%%%%%%%%%%%%%%%%%%%%%%
\sect{Identification of the relaxation spectrum}
It is very easy to identify the perturbed $\lambda_{q,0}$ branch of the spectrum if $\gamma$ is large, because large $\gamma$ shifts all the $\lambda_{q,s{\neq}0}$ eigenvalues to ${\re(\lambda)\sim 2\nu + \gamma}$. But if, say, $\gamma{=}0$, we can still try to identify this branch   
by calculating the diagonal norm $Q$ of each eigen-mode. 
A given eigen-mode $\rho$ of the super-matrix $\mathcal{L}$ can be regarded as a super-vector, with the ad-hoc normalization ${\sum_{x,x'} |\rho_{x,x'}|^2 = 1}$. What we call diagonal norm is the partial sum ${Q=\trc(\rho)}$. In \Fig{fdiagNorm} we demonstrate that in the range of interest this procedure allows to isolate the $\lambda_{q,0}$ branch, even if $\gamma{=}0$.  The points are color coded by the inverse participation ratio, namely, 
\rmrk{${\text{IPR} = \sum_{x,x'} |\rho_{x,x'}|^4}$}. Large IPR for a relaxation-mode indicates localization (only small number of sites participate).

%%%%%%%%%%%%%%%%%%%%%%%%%%%%%%%%%%%%%%%%%%%%%%%%%%%%%%%
\section{Effective disorder}

In order to understand analytically the observed dependence of the spectrum on~$c$, 
we write the equation ${\mathcal{L}\rho{=}-\lambda \rho}$ for the elements ${(x, \ r=-1,0,1)}$ of $\rho$,
based on the diagram of \Fig{fBloch}. Then we eliminate the $r{=}\pm 1$ elements, 
expressing them in terms of the $r{=}0$ elements. Substitution into the equation for the 
$r{=}0$ elements, we conclude that the effective transition rates are modified as follows:
\beq \label{eEff}
\nu_x^{\text{effective}} \ = \ \nu_x + \frac{c^2}{2} \frac{(\lambda-\gamma_x)}{(\lambda-\gamma_x)^2 + \mathcal{E}_x^2} 
\eeq      
This formula allows to estimate how different eigenvalues along the $\lambda_{q,0}$ branch are affected by the disorder.
Let us start our reasoning with the assumption that the bias is small or even zero. Accordingly, the relaxation spectrum is real. 
\Eq{eEff} implies that that the introduction of~$c$ is equivalent to an effective resistor-network disorder with dispersion $\sigma_{\nu}$ that is proportional to $\sigma_{\mathcal{E}}$. For the purpose of rough estimate one can substituted ${ (\lambda{-}\gamma_x) \mapsto \nu }$. Then it follows that  
\beq
\sigma_{\nu}  
% \sim  \ c^2\frac{\nu \mathcal{E}_{\text{bias}}}{(\nu^2+\mathcal{E}_{\text{bias}}^2)^2} \sigma_{\mathcal{E}} 
\ \sim \  \frac{c^2}{\nu^3}T_{\text{bath}}^2 \ \sigma_f^2 
\eeq
where we used ${ \mathcal{E}_{\text{bias}} \ll \sigma_{\mathcal{E}} \ll \nu }$, and the relation \Eq{eT}.  

%%%%%%%%%%%%%%%%%%%%%%%%%%%%%%
\begin{figure}
\centering
\includegraphics[width=4cm]{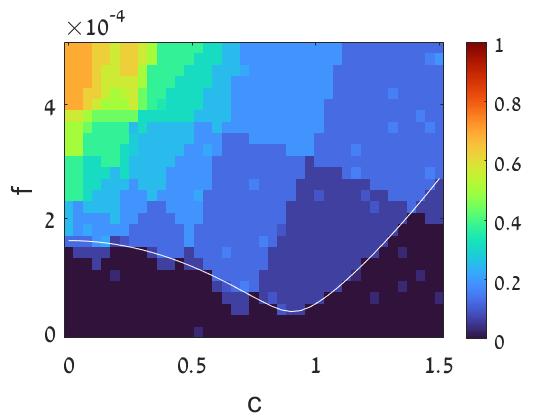}
\includegraphics[width=4cm]{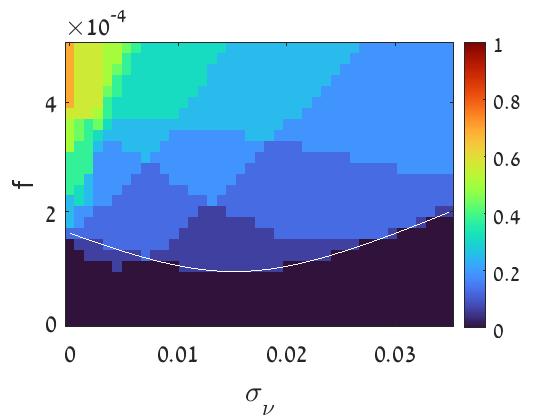}
\caption{
{\bf Enhanced effective disorder.}    
We consider the sample of \Fig{fNcmplx} with \rmrk{${\sigma_f=0.01}$}.
{\em Left panel}:
The relaxation spectrum is obtained from the diagonalization 
of the Lindbladian~$\mathcal{L}$, with \rmrk{${\sigma_{\nu}=0}$}. 
The bath temperature is \rmrk{${ T_{\text{bath}}=800 }$},  
and the on-site decoherence rate is \rmrk{$\gamma=10$}.
The color-code indicates the results for $N_{\text{cmplx}}/N$, 
and the axes are ${(c,f)}$.   
{\em Right panel}:
The results are based on diagonalization of 
the associated  $\mathcal{W}$, 
and the axes are ${(\sigma_{\nu},f)}$. 
Comparing with the left panel one observes qualitative correspondence.  
Note that different samples exhibit different dependence 
on the strength of the effective disorder. 
Monotonic dependence is found only after statistical averaging.  
}
\label{fNcmplxIMGs}
\end{figure}
%%%%%%%%%%%%%%%%%%%%%%%%%%%%%%

%%%%%%%%%%%%%%%%%%%%%%%%%%
\sect{Localization}
Due to $\sigma_f$ and $\sigma_{\nu}$ the eigen-modes of the chain are localized. As the bias is increased gradually from zero, one expects a delocalization transition. We shall discuss this transition analytically in the next section.
We can also go in the other direction. Namely, we fix a relatively large bias, 
such that the relaxation eigen-modes are delocalized, with complex eignevalues $\lambda_q$. 
Then we increase the disorder and/or $c$ gradually, to see how the spectrum is affected. 
We discuss this scenario further below.

In the absence of disorder the introduction of $c$ leads to monotonic increase of $\im(\lambda_q)$, 
as implied by \Eq{eSpec}. This effect is not uniform: 
the eigenvalues in the vicinity of  $\lambda\sim 0$ are hardly affected. 

In the presence of weak disorder the dependence of $\im(\lambda_q)$ on $c$ becomes non-monotonic, see \Fig{fSpctraND}, reflecting a crossover from a non-disordered-like dependence that is implied by \Eq{eSpec} to the disordered-case dependence that is implied by \Eq{eEff}. Namely, the implication of the effective disorder is to ``push" towards localization, hence $\im(\lambda_q)$ is decreased. 

If the effective-disorder is strong enough, the eigenvalues become real, indicating localization. Also here  the effect is not uniform:  the eigenvalues in the vicinity of  $\lambda\sim 0$ are hardly affected. We explain this observation in the next section.

%%%%%%%%%%%%%%%%%%%%%%%%%%%%%%%%%%%%%%%%%%%%%%%%%%
\sect{Global localization}
The global localization of the relaxation modes as a function of $\sigma_f$ for different values of $f$ has been illustrated in lower panel of \Fig{fNcmplx}. 
In the left panel of \Fig{fNcmplxIMGs} we demonstration how this localization is affected by~$c$. We also demonstrate there (in the right panel) that the effect of $c$ can be mimicked by introducing into $\mathcal{W}$ an effective resistor-network-disorder. However, this should not be over-stated. It should be clear that the details of the crossover from non-disordered ring  cannot be captured by a purely stochastic model, because the former features a non-monotonic dependence of the eigenvalues on~$c$.

%%%%%%%%%%%%%%%%%%%%%%%%%%%%%%
\begin{figure}
\centering
\includegraphics[width=4cm]{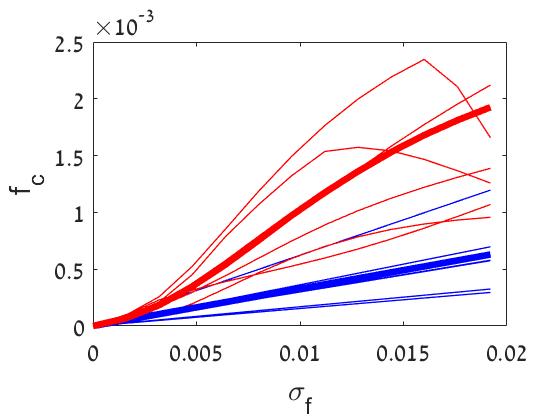}
\includegraphics[width=4cm]{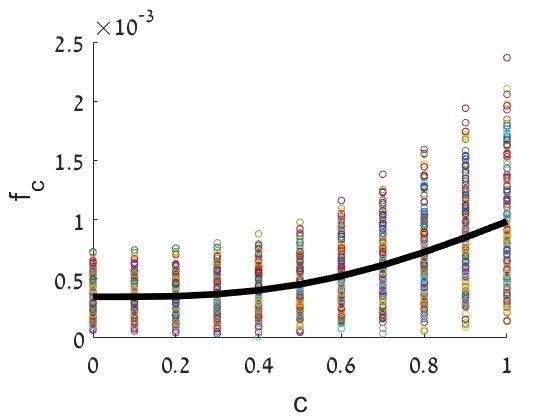} 
\caption{
{\bf Delocalization threshold versus model parameters.}    
The left panel, $f_c$ versus $\sigma_f$,  
is obtained by averaging over 150 realizations of disorder    
for ${c{=}0}$ (thick blue line) and ${c{=}1}$ (thick red line). 
Thin lines illustrate the non-averaged results 
for 6 randomly selected realizations.   
The right panel, $f_c$ versus $c$, 
is for realizations with \rmrk{${\sigma_f=0.01}$}. 
The symbols illustrate the values before averaging.
The bath temperature is \rmrk{${T_{\text{bath}}=200}$},
and the on-site decoherence is \rmrk{${\gamma=0}$}.   
}
\label{fcVSc}
%\end{figure}
%%%%%%%%%%%%%%%%%%%%%%%%%%%%%%
%
%
%
%%%%%%%%%%%%%%%%%%%%%%%%%%%%%%
%\begin{figure}
\centering
\includegraphics[width=4cm]{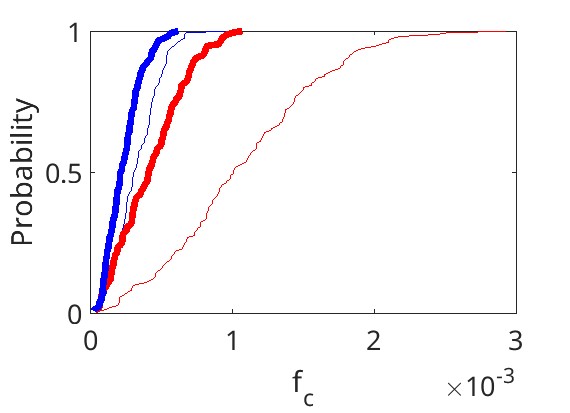} 
\includegraphics[width=4cm]{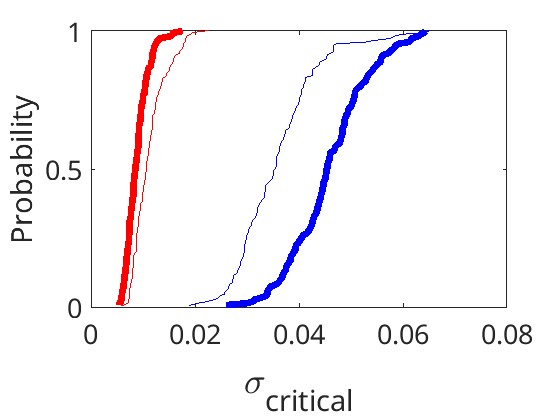} 
\caption{
{\bf Delocalization threshold statistics.}    
Cumulative histograms for $f_c$ given \rmrk{$\sigma_f=0.01$},  
and for $\sigma_{\text{critical}}$ given \rmrk{$f_\text{bias}=0.003$}. 
The thick lines are for ${N=64}$, 
while the thin lines are for ${N=32}$. 
The blue and the red lines are for ${c=0}$ 
and for ${c=1}$ respectively. 
The histograms are based on 150 realizations of the disorder.   
}
\label{fHistog}
\end{figure}
%%%%%%%%%%%%%%%%%%%%%%%%%%%%%%

The delocalization threshold $f_c$ is related to the eigenvalues that reside at the vicinity of $\lambda{=}0$, while the global count $N_{\text{cmplx}}$ of complex eigenvalues probes the delocalization globally.
For the particular disorder-realization of \Fig{fNcmplxIMGs} the dependence on the strength of the disorder is rather monotonic for $N_{\text{cmplx}}$, but not monotonic for $f_c$. For other disorder-realizations the dependence of $f_c$ on the strength of disorder is different. It is only after averaging, over many realizations, that we get a monotonic dependence. Furthermore, we clarify below that the dependence of $f_c$ on $c$ is diminished for large rings.  

The average dependence of $f_c$ on $\sigma_f$ and $c$ is illustrated in \Fig{fcVSc}. We also provide the results for a few randomly selected realizations (thin lines) to illustrate the fluctuations. \Fig{fHistog} displays the full histograms for the 150 disorder-realizations. We see that for larger rings the effect of $c$ on $f_c$ is diminished. This observation will be explained in the next section.

On the other hand the {\em global} effect of $c$ is not diminished for large $N$. For that we have to look globally on the spectrum, and not just at the vicinity of  ${\lambda \sim 0}$. To quantify this statement we find for a given $f_{\text{bias}}$ the critical value $\sigma_{\text{critical}}$ of $\sigma_f$ above which real eigenvalues appear, indicating localization of some `remote' eigen-modes. One observes in the right histogram of \Fig{fHistog} that the effect of $c$ does not diminish for longer samples.

%%%%%%%%%%%%%%%%%%%%%%%%%%%%%%%%%%%%%%%%%%%%%%%%%%%%%%%%%%%%%%%%%%%%%%%%%%%%%%%%%%%%%%%%%%%%%%%%%%%%%%%%%%%%%%%%%%%%
\section{The delocalization threshold}

The localization of the relaxation modes is due to the $\sigma_f$ disorder, and also influenced by the $\sigma_{\nu}$ disorder. The latter is enhanced once  coherent hopping is introduced, as implied by \Eq{eEff}. 

For the purpose of analysis one introduces an Hermitian matrix $\mathcal{H}_{W}$ that is associated with $\mathcal{W}$. Using the same notations as in \Eq{eWmatrix} the associated matrix is 
\beq 
\mathcal{H}_W &=& \text{diagonal}\left\{ (\nu_{x}{+}\nu_{x{-}1}) + \frac{\nu}{2}(f_{x}{-}f_{x{-}1}) + \frac{\nu}{4} f^2 \right\} 
\nonumber \\ \label{eHW}
&& - \text{offdiagonal}\left\{ \nu_x \right\}  
\eeq 
The following relation exists \cite{Hatano1996,Hatano1997,Shnerb1998,neg}
\beq
\det(\lambda{+}\mathcal{W}) &=& \det(\lambda{-}\mathcal{H}_W) 
\nonumber \\
&& - 2\left[\cosh\left(\frac{Nf}{2}\right)-1\right] \prod_x (-\nu_x)  
\ \ \ \ \ \ \ \ \ \ 
\eeq 
Consequently, one can write the characteristic equation for the eigenvalues 
as $F(\lambda)=F(0)$, where 
\beq
F(\lambda) \ = \ \frac{1}{N}\sum_{k=0}^{N} \ln \left|\frac{\lambda - \epsilon_k}{\nu_{\text{avg}}} \right| 
\eeq
Here $\epsilon_k$ are the eigenvalues of $\mathcal{H}_W$, 
and $\nu_{\text{avg}}$ is the geometric average over the $\nu_{x}$. 
The right hand side of the equation $F(\lambda)=F(0)$ is implied 
by the simple observation that $\lambda{=}0$ should 
be a trivial root of the  characteristic equation.

The envelope $\kappa(\lambda)$ of the function $F(\lambda)$ is identified as the Thouless formula for the inverse localization length of eigenstates that are associated with $\epsilon_k$.  Consequently, the condition for getting complex eigenvalues from the equation $F(\lambda)=F(0)$ is $\kappa(\lambda) < F(0)$.  Below we explain the derivation of the following expression
\beq \label{ekappa}
\kappa(\lambda) \ \approx \ 
\alpha_0 f_c - \alpha_c \frac{(f-f_c)}{f_c} \sqrt{\frac{\lambda}{\nu}}
\ + \ \frac{\sigma_{\nu}^2}{8\nu^3} \lambda 
\eeq       
where $f_c$, that is given by \Eq{efc}, is independent of $\sigma_{\nu}$, 
while $\alpha_0$ and $\alpha_c$ are numerical constants.  
From this expression it follows that for ${f>f_c}$ complex roots appear at the vicinity of ${\lambda \sim 0}$. 
This threshold is not affected by $\sigma_{\nu}$, and therefore 
also not affected by $c$.  However, the resistor network disorder 
enhances the localization for larger $\lambda$, and therefore affects 
the {\em global} delocalization of the eigen-modes.        

The derivation of \Eq{ekappa} requires the integration of several ingredients 
that have been worked out in past studies. We provide here an outline how to obtain 
this formula. The basic observation of Derrida and followers is that $\mathcal{W}$ 
generates anaomalous spreading ${|x| \sim t^{\mu} }$ that is characterized by an
exponent $\mu$.  This exponent is determined through the equation 
${ \braket{ e^{-\mu f_x} } = 1 }$. It is important 
to realize that $\mu$ is rigorously independent of the resistor-network disorder. 
For Gaussian distribution one obtains 
the relation ${\mu = (2/\sigma_f^2) f  }$.

It is implied by the anomalous spreading, 
that the density of the eigenvalues $\epsilon_k$ 
at the bottom of the `energy' band is $\epsilon^{\mu-1}$.  
This can be used in the Thouless relation to derive 
the result ${F'(\lambda) \approx  (\lambda^{\mu-1}/\nu^{\mu}) \pi \mu \cot(\pi \mu) }$. 
See \cite{neg} for details. It follows that complex eigenvalues appear 
near the origin for ${\mu > \mu_c}$ where ${\mu_c = (1/2)}$. 
The second term in \Eq{ekappa} is obtained after linearization 
of $F'(\lambda)$ around this critical value.

The first and the third terms in \Eq{ekappa} correspond 
to the correlated Anderson diagonal-disorder, 
and to the Debye-resistor-network off-diagonal disorder 
that we have in \Eq{eHW}. Section VII of \cite{rxa}, 
including Appendix~C there, provide a fair
presentation for these two types of disorder. 
The estimation of the inverse localization length 
is performed using the Born approximation (Fermi-Golden-Rule).
The Debye disorder provides a term that 
is proportional to $[\text{Var}(\nu_x)]\lambda$. 
This contribution vanishes at the bottom of the energy band as expected. 
The Anderson disorder provides a term that
is proportional to $[\text{Var}(\text{diagaonl})]/\lambda$,
where $\text{Var}(\text{diagaonl})$ is the effective variance 
of the diagonal terms. Here one should notice 
that  $f_{x}{-}f_{x{-}1}$ in \Eq{eHW} features telescopic 
correlations, hence  $\text{Var}(\text{diagaonl}) \propto [\text{Var}(f)] \lambda$ 
is proportional to $\lambda$. Consequently the Anderson term in \Eq{ekappa} 
is independent of $\lambda$.

%%%%%%%%%%%%%%%%%%%%%%%%%%%%%%%%%%%%%%%%%%%%%%%%%%%%%%%%%%%%%%%%%%%%%%%%%%%%%%%%%%%%%%%%%%%%%%%%%%%%%%%%%%%%%%%%%%%%
\section{Summary}

Quantum Brownian motion is a well studied theme (see \cite{Schwinger1961,HakimAmbegaokar1985,GRABERT1988115,HanggiQBM2005} and references within). In the condensed-matter literature it is common to refer to the Caldeira-Leggett model \cite{Caldeira1983,CALDEIRA1983374}, where the particle is linearly coupled to the modes of an Ohmic environment. The strongly related  problem of motion in a tight binding lattice \cite{Weiss1985,aslangul1986quantum,AslangulPeriodicPotential1987,Weiss1991} can be regarded as a natural extension of the celebrated spin-boson model. The cited works assume that the fluctuations are uniform in space. Some other works consider the dynamics of a particle that interacts with local baths. In such models the fluctuations acquire finite correlations in space  \cite{dld,MadhukarPost1977,Kumar1985,EspositoGaspard2005,Dibyendu2008,Amir2009,lloyd2011quantum,Moix_2013,CaoSilbeyWu2013,Moix_2013,Kaplan2017ExitSite,Kaplan2017B,qsd}.
More recently, the basic question of transport in a tight-binding lattice has resurfaced 
in the context of excitation transport in photosynthetic light-harvesting complexes    \cite{amerongen2000photosynthetic,ritz2002quantum,plenio2008dephasing,FlemingCheng2009,Rebentrost_2009,Alan2009,Sarovar_2013,higgins2014superabsorption,celardo2012superradiance,park2016enhanced}.

Considering the possibility that each site and each bond experiences a different local bath, it is puzzling  that all of the above cited works have somehow avoided the confrontation of themes that are familiar from the study of stochastic motion in random environment.
Specifically we refer here to the extensive work by Sinai, Derrida, and followers \cite{Sinai1983,Derrida1983,BOUCHAUD1990,BOUCHAUD1990a}, 
and the studies of stochastic relaxation \cite{neg,nep} which is related to the works of Hatano, Nelson and followers
\cite{Hatano1996,Hatano1997,Shnerb1998,Feinberg1999,KAFRI2004,KAFRI2005}.

In order to bridge this gap, we have introduced in an earlier paper \cite{qss} a quantized version of the Sinai-Derrida model. In the present work we have considered a simplified version that captures the essential physics. Without coherent hopping ($c=0$) it reduces to the Pauli master equation, and hence becomes identical to the standard Sinai-Derrida model.  The model features two dimensionless parameters that control its regime diagram \Fig{fRegimes}.  The smallness of the quantum parameter ${\eta \ll 1}$ implies that memory effects can be neglected, hence we can use a Lindblad version of the Ohmic Master equation.  The smallness of the classical parameter ${\sigma_f \ll 1}$ reflects the standard assumption of weak stochastic disorder, as in the original model of Sinai.

The Hatano-Nelson delocalization transition is related to the Sinai-Derrida sliding transition. 
We find that adding coherent transitions ``in parallel" to the stochastic transitions leads to some counter intuitive effects.
In order to illustrate these effects we have inspected mainly two measures: 
{\bf (a)}~the overall number of complex (under-damped) relaxation modes.
{\bf (b)}~the threshold $f_c$ for under-damped relaxation.  
The latter measure focuses on the eigenvalues at the vicinity of ${\lambda \sim 0 }$. 
The main observations are:
{\bf (1)}~The relaxation modes are strongly affected by coherent hopping;  
{\bf (2)}~The dependence of $\im{\lambda}$ on~$c$ becomes non-monotonic.
{\bf (3)}~On-site decoherence affects the sensitivity to the $c$~dependence.
{\bf (4)}~Some features of the localization transition can be mimicked by introducing an effective random-resistor network disorder in the stochastic description. 
{\bf (5)}~The dependence of the delocalization threshold on~$c$ is very weak for large rings. 
{\bf (6)}~The delocalization threshold for small quantum rings exhibits strong fluctuations.

Our observations regarding delocalization concern the regime ${c < \nu}$, within the region where the coherent hopping can be regarded as a perturbation. This means that the relaxation modes are distinct, and well separated form the decoherence modes. This allows a meaningful comparison with the stochastic model.

%%%%%%%%%%%%%%%%%%%%%%%%%%%%%%%%%%%%%%%%%%%%%%%%%%%%%%%%%%%%%%%%%%%%%%%%%%%%%%%%%%%%%%%%%%
%%%%%%%%%%%%%%%%%%%%%%%%%%%%%%%%%%%%%%%%%%%%%%%%%%%%%%%%%%%%%%%%%%%%%%%%%%%%%%%%%%%%%%%%%%
%%%%%%%%%%%%%%%%%%%%%%%%%%%%%%%%%%%%%%%%%%%%%%%%%%%%%%%%%%%%%%%%%%%%%%%%%%%%%%%%%%%%%%%%%%
\appendix
\onecolumngrid

%%%%%%%%%%%%%%%%%%%%%%%%%%%%%%%%%%%%%%%%%%%%%%%%%%%%%%%%%%%%%%%%%%%%%%%%%
\section{The Lindblad master equation}

A master equation for the time evolution of the system probability matrix $\rho$  
is of Lindblad form if it can be written as
\beq \nonumber
\frac{d\rho}{dt} 
\ \ &=& \ \  
-i[\bm{H},\rho] \ + \sum_{x} \nu_{x}  
\left[ \bm{L}_{x} \rho \bm{L}_{x}^{\dagger} \ - \frac{1}{2} \{ \bm{L}_{x}^{\dagger} \bm{L}_{x}, \rho \}  \right]  
\\ \label{eqGenLindblad}
\ \ &=& \ \ 
-i[\bm{H},\rho] \ 
- \frac{1}{2} \left(\mathbf{\Gamma} \rho + \rho \mathbf{\Gamma} \right) 
+ \sum_{x} \nu_{x}  \bm{L}_{x} \rho \bm{L}_{x}^{\dagger} 
\eeq
Here we consider interaction with {\em local} baths that are coupled to the sites or to the bonds, 
hence the index $x$ indicates position. We have defined 
\beq\label{eqGamma}
\bm{\Gamma} = \sum_{x} \nu_x \bm{L}_{x}^{\dagger}\bm{L}_{x}
\eeq 
A Lindblad generator due to coupling to an an Ohmic bath can be written as  
\beq \label{LindbladGen}
\bm{L} \ = \ \bm{W} + i \frac{\eta}{2\nu} \bm{V}  
\eeq
where it has been assumed that the coupling to bath coordinate is $-\bm{W} F_{\text{bath}}$. 
The fluctations of $F$ are charaterized by intensity $\nu$ ("noise") 
and asymmetry $\eta$ ("friction"), while ${ \bm{V} \equiv i[\bm{H},\bm{W}] }$. 
Note that in the Fokker Planck equation $\bm{W}$ is the position coordinate, and $\bm{V}$ is the velocity operator.

%%%%%%%%%%%%%%%%%%%%%%%%
\sect{Bond dissipators}
The interaction with a bath-source that induces non-coherent transitions 
at a given bond is obtained by the replacement ${(c/2)\mapsto (c/2)+ f(t)}$ in 
the respective term of the Hamiltonian.  Accordingly,  
\beq
\bm{W}^{(B)}_x &=& \left(\bm{D}_x + \bm{D}_x^{\dag} \right) \\
\bm{V}^{(B)}_x &=& i[\bm{H}, \bm{W}_x] \ = \
i \mathcal{E}_x \left(\bm{D}_x^{\dag} - \bm{D}_x\right) - 
i \dfrac{c}{2} \left[   \left( \bm{D}_{x+1} \bm{D}_x - \bm{D}_x \bm{D}_{x-1} \right)  - h.c \right].
\eeq
Neglecting the double hopping term the Lindblad generators \Eq{LindbladGen} are 
\beq \label{eLgen}
\bm{L}^{(B)}_x \ = \  \left( 1 + \frac{f_x}{4} \right)\bm{D}_x + \left( 1 - \frac{f_x}{4} \right) \bm{D}_x^\dagger
\eeq
where
\beq 
f_x \ = \ \frac{2\eta_x \mathcal{E}_x}{\nu_x}  \ \equiv \ \frac{\mathcal{E}_x}{T_x}  
\eeq
We identify the rates of transitions 
\beq \label{eWrates}
w^{\pm}_x  \ = \  \left( 1 \pm \frac{f_x}{2} \right) \nu_x  \ = \ \nu_x \pm \eta_x \mathcal{E}_x
\eeq 
Plugging \Eq{eLgen} into \Eq{eqGamma},  
using \Eq{eWrates} and the identity $\bm{D}_x^\dagger \bm{D}_x = \bm{Q}_x$, 
we get,
\beq
\bm{\Gamma}^{(B)} \ \ = \ \ \sum_x  (w^{+}_x +w^{-}_{x{-}1}) \bm{Q}_x \ \ = \ \ (w^{+} +w^{-})
\eeq
where the last expression applies if the rates do not depend on $x$ (no disorder).  
Then from \Eq{eqGenLindblad} we get
\beq
\mathcal{L}^{(B)}\rho = -(w^{+}+w^{-})\rho + \sum_{x}  
\left[
w^{+} \bm{D}_{x}^{\dag} \rho \bm{D}_{x} +  w^{-} \bm{D}_{x} \rho \bm{D}_{x}^{\dag} 
+ \nu  \bm{D}_{x} \rho \bm{D}_{x} + \nu \bm{D}_{x}^{\dag} \rho \bm{D}_{x}^{\dag}
\right] 
\eeq
More generally, with disorder, we get $\mathcal{L}^{(B)}$ of \Eq{eqL}, 
that has been simplified by dropping the last two terms. The omitted terms 
merely modify the lowest decoherence modes as discussed in \cite{qss}. 

%%%%%%%%%%%%%%%%%%%%%%%%
\sect{Site dissipators}
Optionally we can add terms that reflect fluctuations of the field. 
At a given site it is obtained by the replacement ${ U(\bm{x}) \mapsto U(\bm{x}) + \tilde{f}(t)}$, 
where $\tilde{f}(t)$ represents fluctuations of intensity $\gamma$. 
The implied coupling operators are 
\beq 
\bm{W}^{(S)}_x &=& \bm{Q}_x \\
\bm{V}^{(S)}_x &=& i[\bm{H}, \bm{W}^{(S)}_x] \ = \ i (c/2) \left[ (\bm{D}_{x-1}^{\dag} - \bm{D}_{x-1}) - \left( \bm{D}_x^\dag - \bm{D}_{x} \right)\right].
\eeq
Neglecting the hopping effect the Lindblad generator is $\bm{L}^{(S)}_x = \bm{Q}_x$.
To avoid confusion we use $\gamma_x$ instead of $\nu_x$ for the intensity of the bath induced noise.
Plugging into \Eq{eqGamma} we get
\beq
\bm{\Gamma}^{(S)} \ \ = \ \ \sum_x \gamma_x \bm{Q}_x \ \ = \ \ \gamma
\eeq
where the last expression applies if the $\gamma_x$ do not depend on $x$.

\hide{
%%%%%%%%%%%%%%%%%%%%%%%%%%%%%%
\begin{figure}
\centering
\caption{
{\bf Additional figures for the Appendices.}
Checking misc issues and misc approximations, 
for example how spectrum dependends on $T_{bath}$.  
}
\label{fAppendix}
\end{figure}
%%%%%%%%%%%%%%%%%%%%%%%%%%%%%%
}

%%%%%%%%%%%%%%%%%%%%%%%%%%%%%%%%%%%%%%%%%%%%%%%%%%%%%%%%%%%%%%%%%%%%%%%%%%%%%%%%%%%%%%%%%%%%
%%%%%%%%%%%%%%%%%%%%%%%%%%%%%%%%%%%%%%%%%%%%%%%%%%%%%%%%%%%%%%%%%%%%%%%%%%%%%%%%%%%%%%%%%%%%
%\clearpage

\ \\ \ \\

\centerline{*************}

\sect{Acknowledgements} 
This research was supported by the Israel Science Foundation (Grant No.518/22).

%%%%%%%%%%%%%%%%%%%%%%%%%%%%%%%%%%%

%%%%%%%%%%%%%%%%%%%%%%%%%%%%%%%%%%%%%%%%%%%%%%%%%%%%%%%%%%%%%%%%%%%%%%%%%%%%%%%%%%%%%%%%%%%%
%%%%%%%%%%%%%%%%%%%%%%%%%%%%%%%%%%%%%%%%%%%%%%%%%%%%%%%%%%%%%%%%%%%%%%%%%%%%%%%%%%%%%%%%%%%%
\end{document}